\documentclass[longauth,traditabstract,letter]{aa}
\pdfoutput=1

\usepackage{graphicx}
\usepackage{txfonts}
\usepackage{natbib}
\bibpunct{(}{)}{;}{a}{}{,}
\usepackage[pdftex]{hyperref}
\usepackage{ucs}
\usepackage[utf8x]{inputenc}
\usepackage[T1]{fontenc}
\newcommand{\herschel}{{\it Herschel}}
\newcommand\xiico{\ifmmode{\rm ^{12}CO}\else $^{12}$CO\fi}
\newcommand\xiiico{\ifmmode{\rm ^{13}CO}\else $^{13}$CO\fi}
\newcommand\cxvio{\ifmmode{\rm C^{16}O}\else C$^{16}$O\fi}
\newcommand\cxviio{\ifmmode{\rm C^{17}O}\else C$^{17}$O\fi}
\newcommand\cxviiio{\ifmmode{\rm C^{18}O}\else C$^{18}$O\fi}
\graphicspath{{./figures/}}

\begin{document}

\titlerunning{\herschel{}/HIFI observations of CO in the Martian atmosphere}
\title{First results on Martian carbon monoxide from \herschel{}/HIFI observations%
\thanks{\herschel{} is an ESA space observatory with science instruments
provided by European-led Principal Investigator consortia and with
important participation from NASA.}}

\author{
  P.~Hartogh\inst{\ref{inst1}}
  \and M.~I.~B\l{}\k{e}cka\inst{\ref{inst4}}
  \and C.~Jarchow\inst{\ref{inst1}}
  \and H.~Sagawa \inst{\ref{inst1},\ref{inst28}}
  \and E.~Lellouch\inst{\ref{inst2}}
  \and M.~de~Val-Borro\inst{\ref{inst1}}
  \and M.~Rengel\inst{\ref{inst1}}
  \and A.~S.~Medvedev\inst{\ref{inst1}}
  \and B.~M.~Swinyard\inst{\ref{inst9}}
  \and R.~Moreno\inst{\ref{inst2}}
  \and T.~Cavali\'e\inst{\ref{inst1}}
  \and D.~C.~Lis\inst{\ref{inst3}}
  \and M.~Banaszkiewicz\inst{\ref{inst4}}
  \and D.~Bockel\'ee-Morvan\inst{\ref{inst2}}
  \and J.~Crovisier\inst{\ref{inst2}}
  \and T.~Encrenaz\inst{\ref{inst2}}
  \and M.~K\"uppers\inst{\ref{inst7}}
  \and L.-M.~Lara\inst{\ref{inst8}}
  \and S.~Szutowicz\inst{\ref{inst4}}
  \and B.~Vandenbussche\inst{\ref{inst10}}
  \and F.~Bensch\inst{\ref{inst5}}
  \and E.~A.~Bergin\inst{\ref{inst11}}
  \and F.~Billebaud\inst{\ref{inst25}}
  \and N.~Biver\inst{\ref{inst2}}
  \and G.~A.~Blake\inst{\ref{inst3}}
  \and J.~A.~D.~L.~Blommaert\inst{\ref{inst10}}
  \and J.~Cernicharo\inst{\ref{inst12}}
  \and L.~Decin\inst{\ref{inst10},\ref{inst23}}
  \and P.~Encrenaz\inst{\ref{inst13}}
  \and H.~Feuchtgruber\inst{\ref{inst21}}
  \and T.~Fulton\inst{\ref{inst20}}
  \and T.~de~Graauw\inst{\ref{inst18},\ref{inst24},\ref{inst14}}
  \and E.~Jehin\inst{\ref{inst6}}
  \and M.~Kidger\inst{\ref{inst15}}
  \and R.~Lorente\inst{\ref{inst15}}
  \and D.~A.~Naylor\inst{\ref{inst16}}
  \and G.~Portyankina\inst{\ref{inst19}}
  \and M.~S\'anchez-Portal\inst{\ref{inst15}}
  \and R.~Schieder\inst{\ref{inst17}}
  \and S.~Sidher\inst{\ref{inst9}}
  \and N.~Thomas\inst{\ref{inst19}}
  \and E.~Verdugo\inst{\ref{inst15}}
  \and C.~Waelkens\inst{\ref{inst10}}
  \and A.~Lorenzani\inst{\ref{inst22}}
  \and G.~Tofani\inst{\ref{inst22}}
  \and E.~Natale\inst{\ref{inst22}}
  \and J.~Pearson\inst{\ref{inst26}}
  \and T.~Klein\inst{\ref{inst27}}
  \and C.~Leinz\inst{\ref{inst27}}
  \and R.~G\"usten\inst{\ref{inst27}}
  \and C.~Kramer\inst{\ref{inst17}}
  }

\institute{
  Max-Planck-Institut f\"ur Sonnensystemforschung, 37191
    Katlenburg-Lindau, Germany\label{inst1}
  \and Space Research Centre, Polish Academy of Sciences, Warsaw,
    Poland\label{inst4}
  \and Environmental Sensing \& Network Group, NICT,
  4-2-1 Nukui-kita, Koganei, Tokyo 184-8795, Japan\label{inst28}
  \and LESIA, Observatoire de Paris, 5 place Jules Janssen, 92195
    Meudon, France\label{inst2}
  \and STFC Rutherford Appleton Laboratory, Harwell Innovation
    Campus, Didcot, OX11 0QX, UK\label{inst9}
  \and California Institute of Technology, Pasadena, CA 91125, USA\label{inst3}
  \and Rosetta Science Operations Centre, European Space Astronomy
    Centre, European Space Agency, Spain\label{inst7}
  \and Instituto de Astrof\'isica de Andaluc\'ia (CSIC), Spain\label{inst8}
  \and Instituut voor Sterrenkunde, Katholieke Universiteit Leuven,
    Belgium\label{inst10}
  \and DLR, German Aerospace Centre, Bonn-Oberkassel, Germany\label{inst5}
  \and Astronomy Department, University of Michigan, USA\label{inst11}
  \and Universit\'e de Bordeaux, Laboratoire d'Astrophysique de
  Bordeaux, France\label{inst25}
  \and Laboratory of Molecular Astrophysics, CAB-CSIC, INTA, Spain\label{inst12}
  \and Sterrenkundig Instituut Anton Pannekoek, University of Amsterdam,
    Science Park 904, 1098 Amsterdam, The Netherlands\label{inst23}
  \and LERMA, Observatoire de Paris, France\label{inst13}
  \and Max-Planck-Institut f\"ur extraterrestrische Physik,
    Giessenbachstra\ss e, 85748 Garching, Germany\label{inst21}
  \and Bluesky Spectroscopy, Lethbridge, Canada\label{inst20}
  \and SRON Netherlands Institute for Space Research, Landleven 12, 9747
    AD, Groningen, The Netherlands\label{inst18}
  \and Leiden Observatory, University of Leiden, The Netherlands\label{inst24}
  \and Joint ALMA Observatory, Chile\label{inst14}
  \and Institute d'Astrophysique et de Geophysique, Universit\'e de Li\`ege,
    Belgium\label{inst6}
  \and \herschel{} Science Centre, European Space Astronomy
    Centre, European Space Agency, Spain\label{inst15}
  \and Department of Physics and Astronomy, University of Lethbridge,
    Canada\label{inst16}
  \and Physikalisches Institut, University of Bern, Switzerland\label{inst19}
  \and KOSMA, I. Physik. Institut, Universität zu K\"oln, Z\"ulpicher Str.
    77, D 50937 K\"oln, Germany\label{inst17}
  \and Osservatorio Astrofisico di Arcetri-INAF-Largo E.  Fermi 5,
    50100 Florence, Italy\label{inst22}
  \and Jet Propulsion Laboratory, Caltech, Pasadena, CA 91109,
    USA\label{inst26}
  \and Max-Planck-Institut f\"ur Radioastronomie, Auf dem H\"ugel 69, D53121
    Bonn Germany\label{inst27}
  }

\date{Received May 31, 2010; accepted July 6, 2010}

\abstract{We report on the initial analysis of  \herschel{}/HIFI carbon
monoxide (CO) observations of the Martian atmosphere performed between 11 and
16 April 2010. We selected  the (7--6) rotational transitions of the
isotopes \xiiico{} at 771~GHz and \cxviiio{} and 768~GHz in
order to retrieve the mean vertical profile of temperature and the mean
volume mixing ratio of carbon monoxide. The derived temperature profile
agrees within less than 5~K with general circulation model (GCM) predictions
up to an altitude of 45~km, however, show about 12--15~K lower values at
60~km.  The CO mixing ratio was determined as $980 \pm
150$ ppm, in agreement with the 900 ppm derived from \herschel{}/SPIRE
observations in November 2009.}


\maketitle

\section{Introduction}

Carbon monoxide was first definitely detected spectroscopically in the
Martian atmosphere by \citet{1969ApJ...157L.187K}.  From observations
performed in 1967 between Mars $L_s = 110\degr$  and $133\degr$ a volume
mixing ratio (vmr) of 800 ppm was deduced \citet{1977Sci...196.1090K}
reported the first microwave detection of CO performed in 1975, when
observing the J =  1--0 rotational transition during $L_s = 339\degr$.
They found  a vmr of  1900 ppm assuming a disk-averaged continuum
brightness temperature of 200 K and a constant height profile.
\citet{1981Icar...47..166G} observed the same transition in 1980 during
$L_s =  86\degr$ and found 3200 ppm.   The re-analysis of these data by
\citet{1983ApJ...273..829C} resulted in quite different values: $800 \pm
400$ ppm from the \citet{1977Sci...196.1090K} observation and $1400 \pm
500$ ppm from the \citet{1981Icar...47..166G} observation by assuming a
surface brightness temperature of 206 K, a $\pm10$ K variation in model
atmosphere temperatures, and a surface pressure of 6.5 hPa.  They found
2200 ± 1000 ppm, based on their own observations of the CO (2--1) line
performed in 1982 ($L_s = 75\degr$) with the same model assumptions.
\citet{1990JGR....9514543C} derived a mixing ratio of 600 ± 150 ppm from
1988/9 observations ($L_s = 197\degr-341\degr$) of the \xiico{} and
\xiiico{} transitions (1--0 and 2--1) and surface brightness
temperatures between 192 and 216 K. They did not find any significant
temporal variability of CO.  \citet{1990A&A...231L..29R} found a CO vmrs
between 800 and 1200 ppm depending on the orography from Phobos/ISM
observations.  \citet{1991P&SS...39..219L} provided for the first time a
mm-wave map of CO (1988, $L_s = 270\degr$) and deduced no significant
variation from their derived value of 800 ± 200 ppm over the Mars disk.
While the 230 GHz observations provided reasonable surface brightness
temperatures, the values in the 115 GHz band are too high (270 K), very likely
due to sideband ratio calibration errors.  \citet{1998A&A...333.1092B}
retrieved values between 450 and 1150 ppm from ground-based IR
observations in 1990/1.  Also from ground-based  IR observations,
\citet{2003Icar..165..315K} found a hemispheric asymmetry with the CO
vmr increasing from 850 ppm at 23° N to 1250 ppm at 50° S (at $L_s =
112\degr$). Follow-up observations of \citet{2007Icar..190...93K}
between 80° N to 75° S over 4 seasons show even larger variations
between 700 and 1600 ppm.  \citet{2006A&A...459..265E} report about a
seasonal variation in CO over Hellas by a factor of 2 from 2.3 µm CO
band observations by OMEGA/MEX. Infrared observations of PFS/MEX result
in similar variabilities over latitude and season, i.e.\ with a mean
value of 1100 ppm \citep{2009P&SS...57.1446B}.  Finally,
\citet{2009JGRE..11400D03S} derived seasonally and globally averaged 700
ppm CO vmr from CRISM/MRO observations, but with strong seasonal
variations at high latitudes. The summertime near-polar CO mixing ratio
was observed to fall to 200\,ppm in the south and 400\,ppm in the north
during the time the carbon dioxide was sublimating from the seasonal ice
caps.

The determination of the vertical profile of temperature and CO
(together with studies of the carbon and oxygen isotopic ratios in CO)
over different solar longitudes of Mars is one of the goals of the
\herschel{} key programme "Water and related chemistry in the solar
system" \citep{2009P&SS...57.1596H}.  Here we present first results of
observations performed in April 2010 around $L_s = 76\degr$. In this
paper we derive the CO mixing ratio and temperature profile from HIFI
observations.  The latter is also required for analysing other
species not treated in this paper. For simplification we analyse two CO
lines (the 7--6 transitions of the isotopes \xiiico{}  and \cxviiio{})
observed at the same time and in the same sideband.

\section{\herschel{}/HIFI observations}

\begin{figure}
  \centering
  \includegraphics[width=0.5\textwidth]{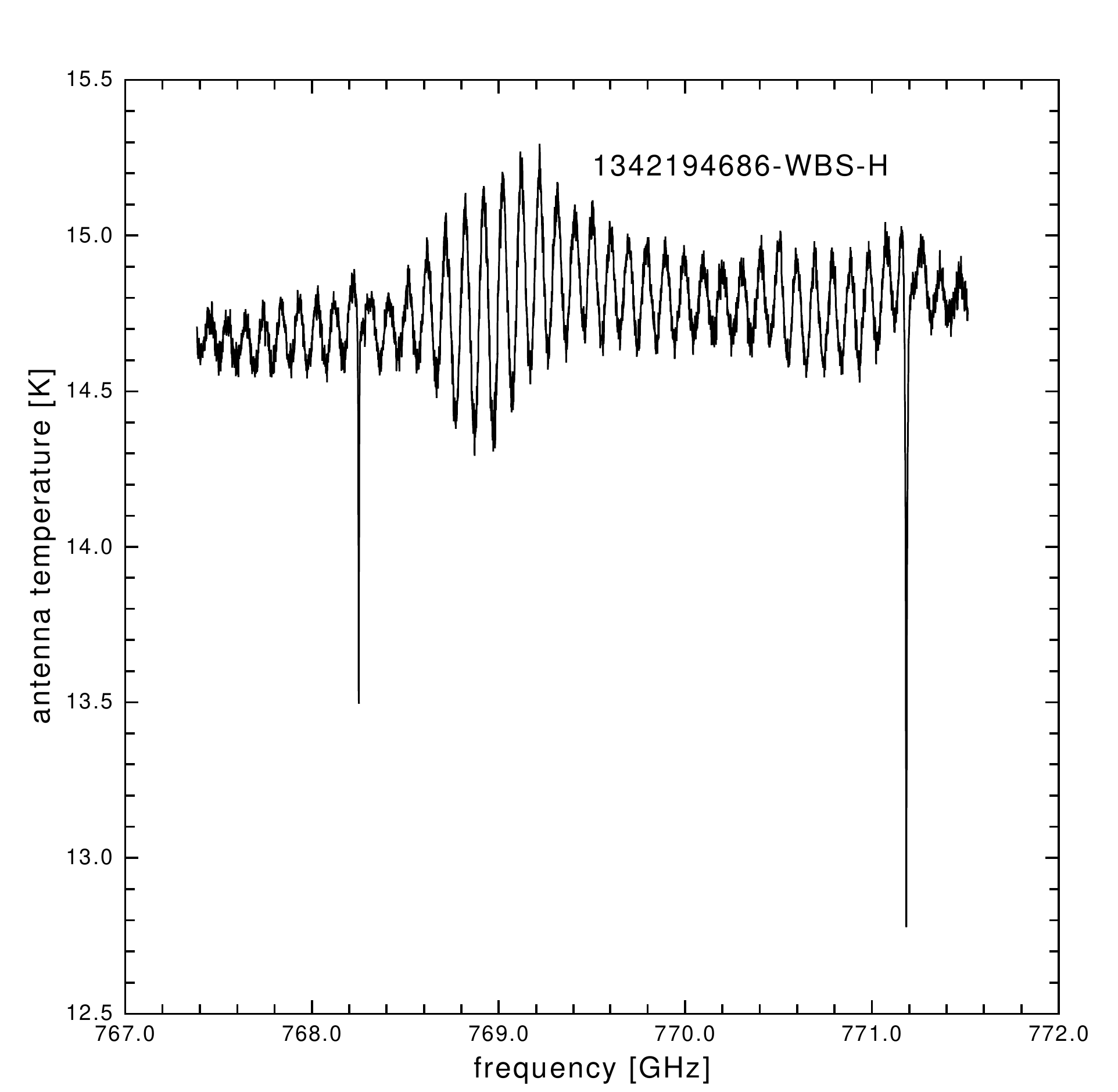}
  \caption{\cxviiio{} (left) and \xiiico{} (right) WBS spectra before
  baseline ripple removal. Both lines are in the same sideband.}
  \label{fig:spect}
\end{figure}

All observations were carried out in HIFI's dual beam switch mode
\citep{2010HIFI,2010Roelfsema} on operational days (OD) 332, 333, 334,
and 337, corresponding to 11--16 April 2010 or $L_s = 75.8\degr$ to 78°.
Dedicated line observations include CO (5--4, 6--5, 8--7), \xiiico{} (7--6),
\cxviio{} (7--6), and \cxviiio{} (7--6). 
HIFI spectrometers can resolve line shapes
with a spectral resolution of 140 KHz (High Resolution Spectrometer; HRS) or
1.1 MHz (Wide Band Spectrometer; WBS).
The integration times range from
93 s for the strongest \xiico{} lines to 9289 s for the \cxviio{}
observation.  
Numerous other CO-lines were detected in the band scans or
as side products of other dedicated line observations, appearing in the
intermediate frequency either in the same or other sideband (double
sideband conversion), some of them with excellent signal-to-noise
ratios. The first set of data was available about a week after the
observations and was processed with the standard HIPE v3.0.1 modules
\citep{2010HIPE} up to level 2. This data set was not complete yet; for
instance, the data of the HRS was only
partly available, and pointing products therein had no entries.  
The
flux calibration by the \herschel{} Science Centre (HSC) was still in
progress and some calibration errors occurred in this first data set.
Furthermore, it turned out in the band scan data that the line amplitudes
of the detected CO lines  were not always exactly the same in the upper
and lower sidebands. Last but not least, the HIFI spectra suffered from a
relatively large baseline ripple (Fig.~\ref{fig:spect}). Fortunately, the
frequency of the ripple generally is the same in the vertical and
horizontal polarizations; however, phase and amplitudes are different 
so they need a dedicated treatment before both polarizations can be
averaged.  Solutions of these problems are under way, but only partly
available now.  Therefore we decided to analyse the data with the
following boundary conditions: (i) line-to-continuum ratio rather than
absolute fluxes (requires assumptions on the surface brightness), (ii)
only horizontal polarization (hence decreases the signal-to-noise ratio), (iii)
only lines in the same sideband, (iv) only WBS
spectra. The surface pressure averaged over the visible Mars disk
changes diurnally in the order of 10\% (derived from the European Mars
Climate Data Base, EMCD v4.1,
\citealt{1999JGR...10424155F,1999JGR...10424177L} and our own general
circulation model, \citealt{2005JGRE..11011008H,2007Icar..186...97M}).
Therefore we introduced boundary condition (v) to analyse only lines
observed at the same time.  
The only data that fulfil these criteria are the (7--6) transitions of
\xiiico{} and \cxviiio{}, which  were observed on 13 April 2010 at 05:33
UT (Obs.\ Id.\ 1342194686).
The integration time for this observation was 900 s.

\section{Analysis and discussion}

Compared to cometary observations of HIFI
\citep{2010arXiv1005.2969H,2010Wild}, the baseline ripple on the Mars
observations is rather large because of its strong continuum emission.
This phenomenon was frequently observed by ground-based observations of
planets. While in the cometary case, the baseline ripple has been removed
with a polynomial fit, for Mars we determined the baseline
frequencies by a normalized periodogram according to
\citet{1976Ap&SS..39..447L} and subtracted 3 periods from the original
spectrum.  
The influence of the line shapes on the determination of the baseline
periods is negligible  when taking the large number of periods
in the spectrum into account.
The observed spectral lines have been modelled using a
standard radiative transfer code: Mars was assumed as a perfect sphere
surrounded by a set of hundred concentric atmospheric layers of 1 km
thickness each \citep[as in][]{2008P&SS...56.1368R}.  Within each layer
the atmospheric temperature, pressure, and volume-mixing ratio of carbon
monoxide have been assumed constant.  The surface continuum emission was
modelled as blackbody emission using a temperature distribution falling
off towards the edge of the apparent disk according to $T(\alpha) = T_0
\times (1 - 0.2\times(1 - \cos(\alpha)))$, with $\alpha$ running from 0°
(nadir) to 90° (limb) across the apparent disk \citep[limb darkening,
compare also][]{2008A&A...489..795C}.

The emission was obtained by integrating over the apparent disk using 64
concentric rings distributed unevenly over the disk and the limb region.
The variation in the path lengths through the atmosphere were 
taken into account fully when calculating the radiation transfer for each
ring. In our model the total continuum flux emitted by the surface
depends purely on the choice of the temperature $T_0$, which defines the
temperature scale for the temperature profile to be retrieved. We have
adjusted $T_0$ in such a way as to exactly match the total flux predicted
by the `Mars continuum model' of $\sim 4230$ Jy provided by
\citet{2008Lellouch} and to match the temperature fall off towards the limb
therein by a factor 0.2.
The error of the modelled flux is 5\%.

Absorption coefficients for the CO spectral lines were calculated
using the HITRAN 2008 spectral line catalogue, keeping the terrestrial
isotopic ratios in it; i.e., \cxvio{}/\cxviiio{} = 498.70 and
\xiico{}/\xiiico{} = 89.01.  \citet{2007Icar..192..396K} conclude that
the deviations of these isotopic ratios on Mars compared to Earth are
less than 2\%.  However, to account for carbon dioxide instead of air as
broadening gas, the broadening parameters provided by the catalogue were 
multiplied by a factor of 1.4 according to
\citet{1982JQSRT..28..409N}.

\begin{figure*}
  \centering
  \includegraphics[width=.92\textwidth]{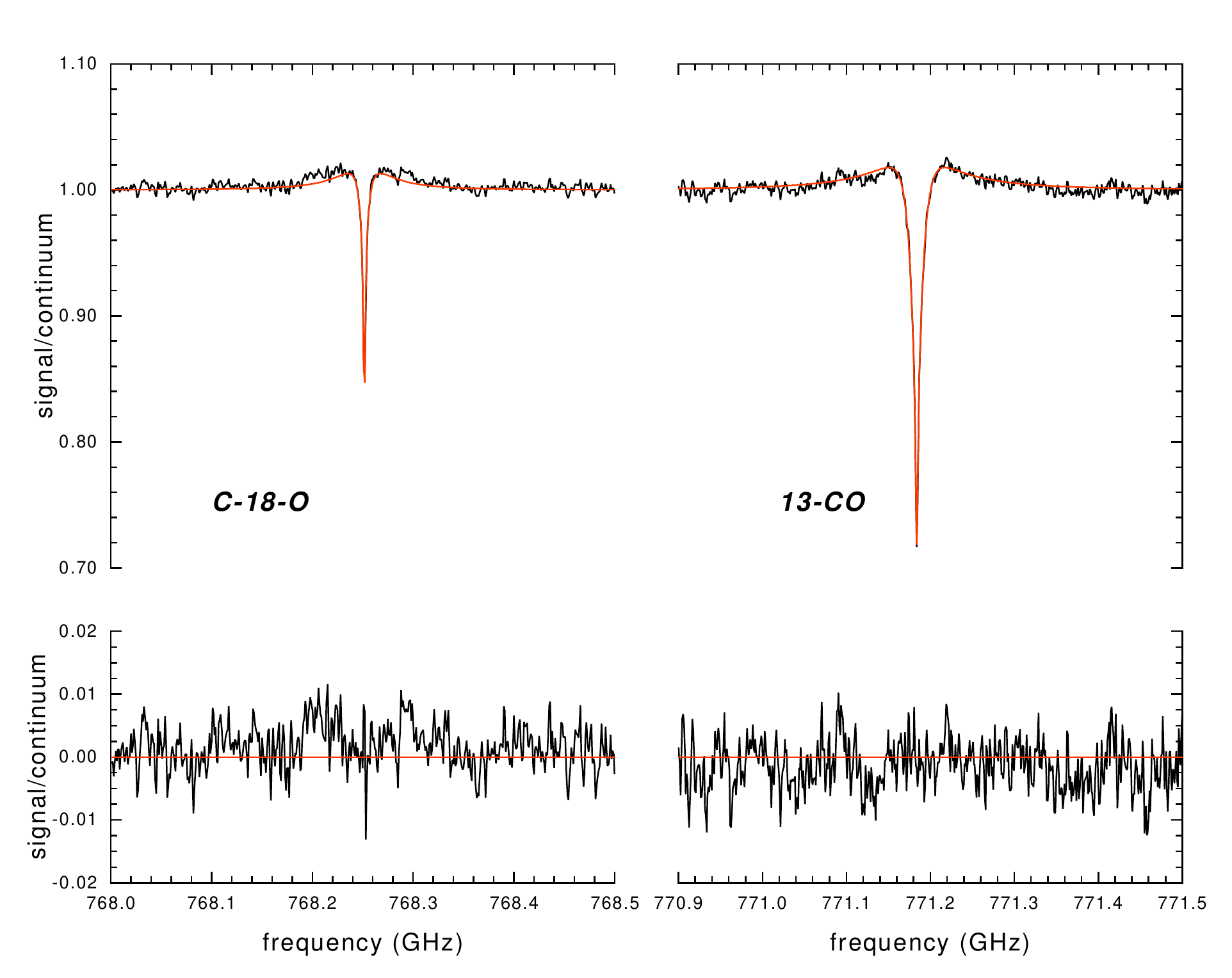}
  \caption{The \xiiico{} and \cxviiio{} lines after removal of the baseline
  ripples. A radiative transfer model was fitted simultaneously to the
  spectra to retrieve temperature profile and volume mixing
  ratio of CO.  The lower panels show the differences between model and
  observations.}
  \label{fig:model}
\end{figure*}

\begin{figure*}
  \centering
  \includegraphics[width=.93\textwidth]{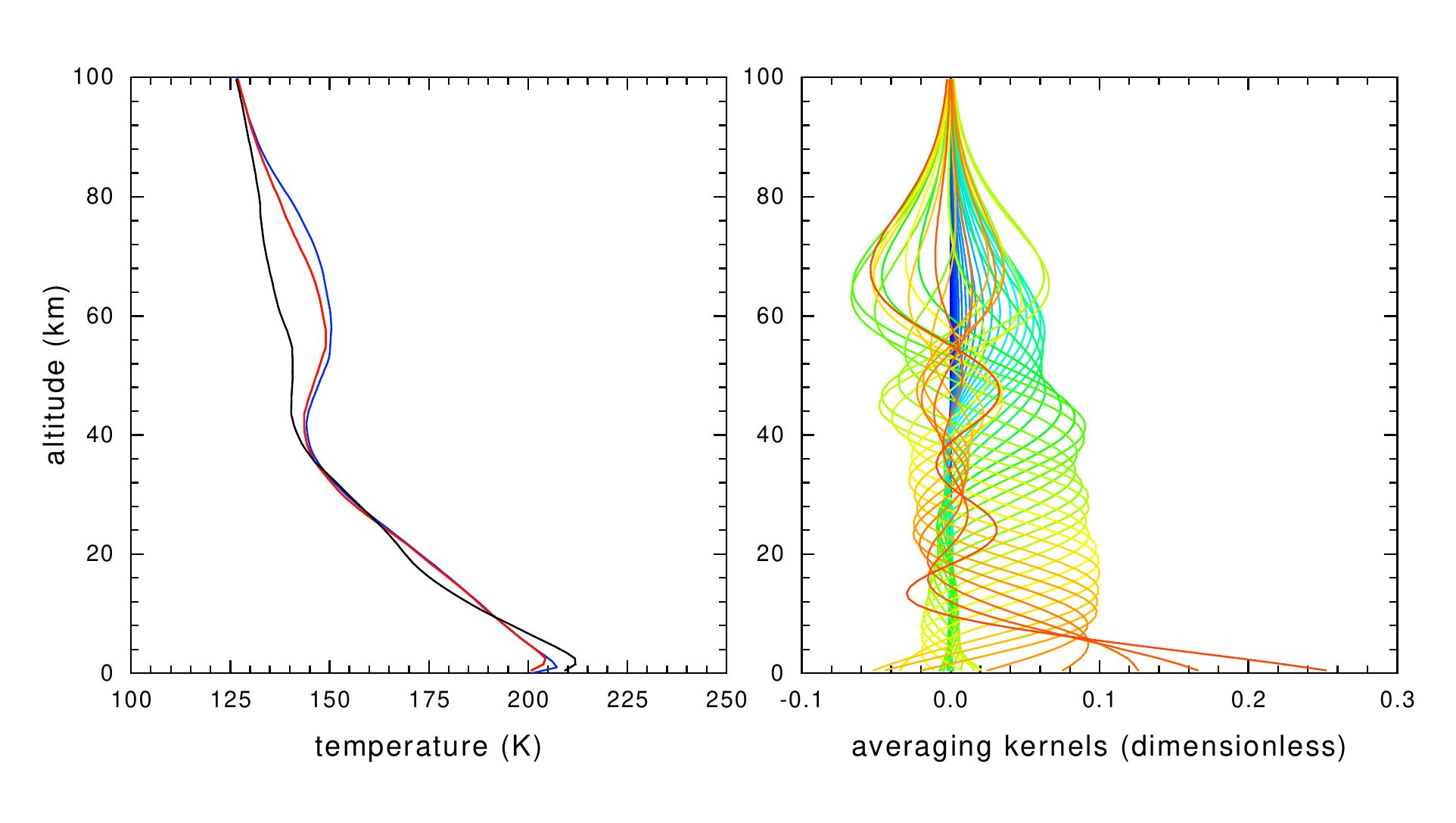}
  \caption{Retrieved vertical profile of temperature (left) and
  averaging kernels (right).  The averaging kernels express the response
  of the fitted temperature profile to a delta function perturbation in
  the true profile at a certain altitude \citep[see][for a sound
  introduction of this quantity]{1990JGR....95.5587R}.  Different colours
  are used to distinguish the averaging kernels belonging to
  perturbations at the different altitude levels.  
  The blue temperature profile is
  derived from EMCD.  
  }
  \label{fig:retrieval}
\end{figure*}

To obtain a temperature profile and the mixing ratio of  CO  we
employed Rodger's optimal estimation method
\citep{1976RvGeo..14..609R,1990JGR....95.5587R}.  This method uses
so-called `a priori' information -- in our case the best estimate of the
temperature profile and CO mixing ratio profile -- which is then updated
to allow for a best fit of the spectral line shapes (minimization of
$\chi^2$). This a priori profile was taken from our MAOAM general
circulation model \citep{2005JGRE..11011008H} considering
the exact observation date, geometry, and time of the HIFI observations.
The surface pressure averaged over the visible disk was 6.7 hPa.
Compared to temperature profiles derived from the EMCD, we find
temperature differences less than 3 K between 3--60 km. EMCD provides about
5 K higher temperatures below 3 km and 3 to 5 K higher temperatures
between 60--80 km. Up to 100 km, both models slowly merge to almost
the same temperature.
It is worth noting a high degree of coincidence of the
averaged temperature profiles over the field of view of the telescope
with the two models. Even though the one from EMCD represents monthly
averaged fields, the one from MAOAM is based on an instantaneous
snapshot, and generally the altitude-latitude distributions differ. This
increases our degree of confidence in the simulated temperatures.

A simultaneous fit of the two spectral lines allows retrieving
temperature and mixing ratios independently, because of the rather
different optical depth of the two lines, where the \xiiico{} line is
optically thick ($\tau = 6.3$ in line centre) and the \cxviiio{} line
being optically thin ($\tau=1.1$).  Figure~\ref{fig:model} shows the
simultaneously fitted spectra of \xiiico{} and \cxviiio{} (and the
residuals) after the removal of the baseline ripple. 
The retrieved CO mixing
ratio  amounts to $980 \pm 100$ ppm which
is in agreement with the one detected by SPIRE observations during $L_s
= 5\degr$ from 6 November 2009 at 20:20 UT \citep{2010arXiv1005.4579S}
under rather similar surface pressure conditions. 
The SPIRE value of 900 ppm has been used as an a priori input to the
retrieval algorithm.
Figure~\ref{fig:retrieval} shows the corresponding temperature profile
and the averaging kernels. The latter provide information about the
sensitivity of the retrieval versus altitude. 
Although the contribution of the a priori profile to the retrieved
temperature profile is less than 10\% below 60 km, it fits quite well to the
profiles predicted by the GCMs. However, the differences are least 
with EMCD near the ground ($\sim 5$ K) and with MAOAM near 60--70 km.
Nevertheless, the observations provide about 12--15 K lower temperatures
near 65 km. The temperature inversion between 40--60 km predicted by
the GCMs should be manifested  in an emission feature in the centre of
the CO lines, which obviously is not the case. 
The 5\% error of the model continuum flux translates into a
roughly 5\% shift of the temperature profiles, i.e., all temperatures will
be about 10 K higher or lower than the retrieved value. In the first
case, the agreement with the model profile is better between 40--80 km
(although still without temperature inversion) but worse below 40 km. In
the second case the agreement is worse for all altitudes.

\section{Summary}

This work presents the first simultaneous retrievals of temperature and
carbon monoxide in the Martian atmosphere derived from HIFI data. The
temperature profile can be used as an input parameter for
determining of concentrations of other gases observed by HIFI during
the same period.  Future work will include all observed CO transitions
in order to better constrain the temperature profile above 60 km and
take advantage of the much wider opacity range for retrieving the
vertical profile of CO.

\begin{acknowledgements}
HIFI has been designed and built by a consortium of institutes and university
departments from across Europe, Canada, and the United States under the
leadership of SRON Netherlands Institute for Space Research, Groningen, The
Netherlands, and with major contributions from Germany, France, and the US.
Consortium members are: Canada: CSA, U.Waterloo; France: CESR, LAB, LERMA,
IRAM; Germany: KOSMA, MPIfR, MPS; Ireland, NUI Maynooth; Italy: ASI, IFSI-INAF,
Osservatorio Astrofisico di Arcetri-INAF; Netherlands: SRON, TUD; Poland: CAMK,
CBK; Spain: Observatorio Astronómico Nacional (IGN), Centro de Astrobiología
(CSIC-INTA). Sweden: Chalmers University of Technology - MC2, RSS \& GARD;
Onsala Space Observatory; Swedish National Space Board, Stockholm University -
Stockholm Observatory; Switzerland: ETH Z\"urich, FHNW; USA: Caltech, JPL, NHSC.
HIPE is a joint development by the \herschel{} Science Ground Segment Consortium,
consisting of ESA, the NASA \herschel{} Science Center, and the HIFI, PACS and
SPIRE consortia.
This development has been supported by national funding agencies: CEA, CNES,
CNRS (France); ASI (Italy); DLR (Germany).
Additional funding support for some instrument activities has been provided by
ESA.
Support for this work was provided by NASA through an award issued by
JPL/Caltech.
MIB, MB, and SS are supported by the Polish Ministry of Education and Science
(MNiSW).
DCL is supported by the NSF, award AST-0540882 to the Caltech
Submillimeter Observatory.
\end{acknowledgements}

\bibliographystyle{aa}
\bibliography{ads,preprints}

\begin{thebibliography}{32}
\expandafter\ifx\csname natexlab\endcsname\relax\def\natexlab#1{#1}\fi

\bibitem[{{Billebaud} {et~al.}(2009){Billebaud}, {Brillet}, {Lellouch},
  {Fouchet}, {Encrenaz}, {Cottini}, {Ignatiev}, {Formisano}, {Giuranna},
  {Maturilli}, \& {Forget}}]{2009P&SS...57.1446B}
{Billebaud}, F., {Brillet}, J., {Lellouch}, E., {et~al.} 2009, \planss, 57,
  1446

\bibitem[{{Billebaud} {et~al.}(1998){Billebaud}, {Rosenqvist}, {Lellouch},
  {Maillard}, {Encrenaz}, \& {Hourdin}}]{1998A&A...333.1092B}
{Billebaud}, F., {Rosenqvist}, J., {Lellouch}, E., {et~al.} 1998, \aap, 333,
  1092

\bibitem[{{Cavali{\'e}} {et~al.}(2008){Cavali{\'e}}, {Billebaud}, {Encrenaz},
  {Dobrijevic}, {Brillet}, {Forget}, \& {Lellouch}}]{2008A&A...489..795C}
{Cavali{\'e}}, T., {Billebaud}, F., {Encrenaz}, T., {et~al.} 2008, \aap, 489,
  795

\bibitem[{{Clancy} \& {Muhleman}(1983)}]{1983ApJ...273..829C}
{Clancy}, R.~T. \& {Muhleman}, D.~O. 1983, \apj, 273, 829

\bibitem[{{Clancy} {et~al.}(1990){Clancy}, {Muhleman}, \&
  {Berge}}]{1990JGR....9514543C}
{Clancy}, R.~T., {Muhleman}, D.~O., \& {Berge}, G.~L. 1990, \jgr, 95, 14543

\bibitem[{{de Graauw} {et~al.}(2010){de Graauw}, {Helmich}, {Philipps}, {}, {},
  \& {}}]{2010HIFI}
{de Graauw}, {\relax Th}., {Helmich}, F.~P., {Philipps}, T.~G., {et~al.} 2010,
  \aap\ in press

\bibitem[{{de Val-Borro} {et~al.}(2010){de Val-Borro}, {Hartogh}, {Crovisier},
  {Bockel{\'e}e-Morvan}, {Biver}, {Lis}, {Moreno}, \& {Jarchow}}]{2010Wild}
{de Val-Borro}, M., {Hartogh}, P., {Crovisier}, J., {et~al.} 2010, \aap\ in
  press

\bibitem[{{Encrenaz} {et~al.}(2006){Encrenaz}, {Fouchet}, {Melchiorri},
  {Drossart}, {Gondet}, {Langevin}, {Bibring}, {Forget}, \&
  {B{\'e}zard}}]{2006A&A...459..265E}
{Encrenaz}, T., {Fouchet}, T., {Melchiorri}, R., {et~al.} 2006, \aap, 459, 265

\bibitem[{{Forget} {et~al.}(1999){Forget}, {Hourdin}, {Fournier}, {Hourdin},
  {Talagrand}, {Collins}, {Lewis}, {Read}, \& {Huot}}]{1999JGR...10424155F}
{Forget}, F., {Hourdin}, F., {Fournier}, R., {et~al.} 1999, \jgr, 104, 24155

\bibitem[{{Good} \& {Schloerb}(1981)}]{1981Icar...47..166G}
{Good}, J. \& {Schloerb}, F.~P. 1981, Icarus, 47, 166

\bibitem[{{Hartogh} {et~al.}(2010){Hartogh}, {Crovisier}, {de Val-Borro},
  {Bockel{\'e}e-Morvan}, {Biver}, {Lis}, {Moreno}, {Jarchow}, {Rengel},
  {Emprechtinger}, {Szutowicz}, {Banaszkiewicz}, {Bensch}, {Blecka},
  {Cavali{\'e}}, {Encrenaz}, {Jehin}, {K{\"u}ppers}, {Lara}, {Lellouch},
  {Swinyard}, {Vandenbussche}, {Bergin}, {Blake}, {Blommaert}, {Cernicharo},
  {Decin}, {Encrenaz}, {de Graauw}, {Hutsemekers}, {Kidger}, {Manfroid},
  {Medvedev}, {Naylor}, {Schieder}, {Thomas}, {Waelkens}, {Roelfsema},
  {Dieleman}, {Guesten}, {Klein}, {Kasemann}, {Caris}, {Olberg}, \&
  {Benz}}]{2010arXiv1005.2969H}
{Hartogh}, P., {Crovisier}, J., {de Val-Borro}, M., {et~al.} 2010, \aap\, in
  press

\bibitem[{{Hartogh} {et~al.}(2009){Hartogh}, {Lellouch}, {Crovisier},
  {Banaszkiewicz}, {Bensch}, {Bergin}, {Billebaud}, {Biver}, {Blake}, {Blecka},
  {Blommaert}, {Bockel{\'e}e-Morvan}, {Cavali{\'e}}, {Cernicharo}, {Courtin},
  {Davis}, {Decin}, {Encrenaz}, {Encrenaz}, {Gonz{\'a}lez}, {de Graauw},
  {Hutsem{\'e}kers}, {Jarchow}, {Jehin}, {Kidger}, {K{\"u}ppers}, {de Lange},
  {Lara}, {Lis}, {Lorente}, {Manfroid}, {Medvedev}, {Moreno}, {Naylor},
  {Orton}, {Portyankina}, {Rengel}, {Sagawa}, {S{\'a}nchez-Portal}, {Schieder},
  {Sidher}, {Stam}, {Swinyard}, {Szutowicz}, {Thomas}, {Thornhill},
  {Vandenbussche}, {Verdugo}, {Waelkens}, \& {Walker}}]{2009P&SS...57.1596H}
{Hartogh}, P., {Lellouch}, E., {Crovisier}, J., {et~al.} 2009, \planss, 57,
  1596

\bibitem[{{Hartogh} {et~al.}(2005){Hartogh}, {Medvedev}, {Kuroda}, {Saito},
  {Villanueva}, {Feofilov}, {Kutepov}, \& {Berger}}]{2005JGRE..11011008H}
{Hartogh}, P., {Medvedev}, A.~S., {Kuroda}, T., {et~al.} 2005, JGR, 110, 11008

\bibitem[{{Kakar} {et~al.}(1977){Kakar}, {Walters}, \&
  {Wilson}}]{1977Sci...196.1090K}
{Kakar}, R.~K., {Walters}, J.~W., \& {Wilson}, W.~J. 1977, Science, 196, 1090

\bibitem[{{Kaplan} {et~al.}(1969){Kaplan}, {Connes}, \&
  {Connes}}]{1969ApJ...157L.187K}
{Kaplan}, L.~D., {Connes}, J., \& {Connes}, P. 1969, \apjl, 157, L187+

\bibitem[{{Krasnopolsky}(2003)}]{2003Icar..165..315K}
{Krasnopolsky}, V.~A. 2003, Icarus, 165, 315

\bibitem[{{Krasnopolsky}(2007)}]{2007Icar..190...93K}
{Krasnopolsky}, V.~A. 2007, Icarus, 190, 93

\bibitem[{{Krasnopolsky} {et~al.}(2007){Krasnopolsky}, {Maillard}, {Owen},
  {Toth}, \& {Smith}}]{2007Icar..192..396K}
{Krasnopolsky}, V.~A., {Maillard}, J.~P., {Owen}, T.~C., {Toth}, R.~A., \&
  {Smith}, M.~D. 2007, Icarus, 192, 396

\bibitem[{Lellouch \& Amri(2008)}]{2008Lellouch}
Lellouch, E. \& Amri, H. 2008,
  \url{http://www.lesia.obspm.fr/perso/emmanuel-lellouch/mars/}

\bibitem[{{Lellouch} {et~al.}(1991){Lellouch}, {Paubert}, \&
  {Encrenaz}}]{1991P&SS...39..219L}
{Lellouch}, E., {Paubert}, G., \& {Encrenaz}, T. 1991, \planss, 39, 219

\bibitem[{{Lewis} {et~al.}(1999){Lewis}, {Collins}, {Read}, {Forget},
  {Hourdin}, {Fournier}, {Hourdin}, {Talagrand}, \&
  {Huot}}]{1999JGR...10424177L}
{Lewis}, S.~R., {Collins}, M., {Read}, P.~L., {et~al.} 1999, \jgr, 104, 24177

\bibitem[{{Lomb}(1976)}]{1976Ap&SS..39..447L}
{Lomb}, N.~R. 1976, \apss, 39, 447

\bibitem[{{Medvedev} \& {Hartogh}(2007)}]{2007Icar..186...97M}
{Medvedev}, A.~S. \& {Hartogh}, P. 2007, Icarus, 186, 97

\bibitem[{{Nakazawa} \& {Tanaka}(1982)}]{1982JQSRT..28..409N}
{Nakazawa}, T. \& {Tanaka}, M. 1982, Journal of Quantitative Spectroscopy and
  Radiative Transfer, 28, 409

\bibitem[{Ott(2010)}]{2010HIPE}
Ott, S. 2010, ASP Conference Series, Astronomical Data Analysis Software and
  Systems XIX, Y. Mizumoto, K.-I. Morita, and M. Ohishi, eds., in press

\bibitem[{{Rengel} {et~al.}(2008){Rengel}, {Hartogh}, \&
  {Jarchow}}]{2008P&SS...56.1368R}
{Rengel}, M., {Hartogh}, P., \& {Jarchow}, C. 2008, \planss, 56, 1368

\bibitem[{{Rodgers}(1976)}]{1976RvGeo..14..609R}
{Rodgers}, C.~D. 1976, Reviews of Geophysics, 14, 609

\bibitem[{{Rodgers}(1990)}]{1990JGR....95.5587R}
{Rodgers}, C.~D. 1990, \jgr, 95, 5587

\bibitem[{{Roelfsema} {et~al.}(2010){Roelfsema}, {Helmich}, {Teyssier}, \& {et
  al.}}]{2010Roelfsema}
{Roelfsema}, P., {Helmich}, F., {Teyssier}, D., \& {et al.} 2010, \aap\ this
  issue

\bibitem[{{Rosenqvist} {et~al.}(1990){Rosenqvist}, {Bibring}, {Combes},
  {Drossart}, {Encrenaz}, {Erard}, {Forni}, {Gondet}, {Langevin}, {Lellouch},
  {Masson}, \& {Soufflot}}]{1990A&A...231L..29R}
{Rosenqvist}, J., {Bibring}, J., {Combes}, M., {et~al.} 1990, \aap, 231, L29

\bibitem[{{Smith} {et~al.}(2009){Smith}, {Wolff}, {Clancy}, \&
  {Murchie}}]{2009JGRE..11400D03S}
{Smith}, M.~D., {Wolff}, M.~J., {Clancy}, R.~T., \& {Murchie}, S.~L. 2009,
  Journal of Geophysical Research (Planets), 114, 0

\bibitem[{{Swinyard} {et~al.}(2010){Swinyard}, {Hartogh}, {Sidher}, {Fulton},
  {Lellouch}, {Jarchow}, {Griffin}, {Moreno}, {Sagawa}, {Portyankina},
  {Blecka}, {Banaszkiewicz}, {Bockelee-Morvan}, {Crovisier}, {Encrenaz},
  {Kueppers}, {Lara}, {Lis}, {Medvedev}, {Renge}, {Szutowicz}, {Vandenbussche},
  {Bensch}, {Bergin}, {Billebaud}, {Biver}, {Blake}, {Blommaert}, {de
  Val-Borro}, {Cernicharo}, {Cavalie}, {Courtin}, {Davis}, {Decin}, {Encrenaz},
  {de Graauw}, {Jehin}, {Kidger}, {Leeks}, {Orton}, {Naylor}, {Schieder},
  {Stam}, {Thomas}, {Verdugo}, {Waelkens}, \& {Walker}}]{2010arXiv1005.4579S}
{Swinyard}, B.~M., {Hartogh}, P., {Sidher}, S., {et~al.} 2010, \aap\ in press

\end{thebibliography}

\end{document}